# 31.1 A 14.08-to-135.69Token/s ReRAM-on-Logic Stacked Outlier-Free Large-Language-Model Accelerator with Block-Clustered Weight-Compression and Adaptive Parallel-Speculative-Decoding

Pingcheng Dong[1,2], Yonghao Tan[1,2], Xuejiao Liu[2], Peng Luo[2], Yu Liu[2], Di Pang[2], Songchen Ma[1,2], Xijie Huang[1], Shih-Yang Liu[1], Dong Zhang[1,2], Zhichao Lu[3], Luhong Liang[2], Chi-Ying Tsui[1,2], Fengbin Tu[1,2], Liang Zhao[4], Kwang-Ting Cheng[1,2]

[1]Hong Kong University of Science and Technology, Hong Kong, China, [2]AI Chip Center for Emerging Smart System, Hong Kong, China, [3]Hefei Reliance Memory, Hefei, China, [4]Zhejiang University, Hangzhou, China

## Abstract

This work presents a 55nm speculative decoding-based LLM accelerator with bumping-based face-to-face ReRAM-on-logic stacking technology. It features a local rotation unit for outlier-free low-bit quantization, a stacking-aware PNM architecture co-designed with blockwise vector quantization to reduce weight EMA overheads, and an adaptive parallel speculative decoding scheme with out-of-order scheduler for high resource and bandwidth utilization. Our chip achieves 14.08-to-135.69token/s and 4.46-to-7.17× speedup over vanilla speculative decoding.

Large Language Models (LLMs) [1-2] have achieved exceptional performance in natural language processing tasks, but their token-by-token autoregressive decoding (AD) paradigm incurs severe latency overhead due to extensive weight external memory access (EMA) [3-7]. Recently, speculative decoding (SD) [8-9] shown in Fig. 31.1.1 has been proposed and widely adopted on GPUs to resolve this issue by using a small-scale draft LLM (DLM) to decode multiple tokens in advance, which are then verified in parallel by a large-scale target LLM (TLM). However, on resource-constrained edge accelerators, LLM latency under SD is still dominated by weight EMA, with over 60% stemming from TLM. Although longer draft length (DL) can reduce TLM weight EMA by yielding more accepted tokens, increasing DLM latency diminishes this potential latency improvement. Thus, the TLM and DLM synergistically become the bottlenecks of SD, raising three challenges: 1) Low-bit post-training quantization [10] is frequently employed to reduce TLM EMA, but it suffers from severe accuracy degradation due to activation outliers [11]. Recent post-training quantization (PTQ) methods based on Fast Walsh-Hadamard Transform (FWHT) [11-13] can eliminate outliers, while preserving computational invariance due to the orthogonal property of the Hadamard matrix, but the deep FWHT array required by varied dimensions in TLM occupies nearly 4.37× the area of a 4K INT8 multiply-accumulate (MAC) array, incurring heavy area overheads. 2) Although DLM is much smaller and easily quantized to low bit precision via quantization-aware training (QAT), limited on-chip memory capacity in edge accelerators still fails to buffer all DLM weights, forcing frequent EMA where constrained external bandwidth further exacerbates latency overheads. 3) At long DL, over 90% of draft tokens decoded from DLM are rejected by TLM, whose latency overhead outweighs the latency savings from reduced TLM EMA.

To overcome these challenges, we develop an SD-based LLM accelerator with three key features: 1) We design a local rotation unit (LRU) that approximates global rotation by decomposing the deep FWHT into overlapped upper and lower low-cost 6-depth FWHTs. By rotating token features in two stages, the LRU removes activation outliers with little area burden and attains a 3.82-to-3.93× speedup over vanilla SD. 2) To extend on-chip memory capacity and avoid DLM EMA in a low-power yet cost-effective way, we utilize bumping-based ReRAM-on-logic stacking technology to design a ReRAM-stacked process-near-memory (RS-PNM) architecture with blockwise vector quantization (BVQ) algorithm. BVQ clusters DLM weights into block-level codebooks (CBs) stored in the high-density ReRAM, while RS-PNM reconstructs weights by retrieving CBs via a high-bandwidth stacking interface, which achieves 1.1-to-1.46× speedup over W4A8 SD with a LRU. 3) Inspired by recent parallel SD [14], we propose an adaptive parallel SD (APSD) scheme that combines the low rejection ratio of short DL and the high accepted token yield of long DL. APSD begins with non-parallel short DL drafting, followed by parallel draft-and-verify. Then, it dynamically adapts the drafting strategy based on feedback from TLM verification. If TLM accepts all previous draft tokens and its newest generated token matches the first draft token from concurrent DLM drafting, the parallel draft-and-verify continues. Otherwise, the draft tokens are discarded and APSD reverts to non-parallel DLM drafting. We further design an out-of-order scheduler with 4 parallel instruction queues decoupled from APSD workloads to avoid resource competition in parallel draft-and-verify as mentioned by [14], which attains 1.1-to-1.29× speedup and 10-to-14% rejected token ratio reduction.

Figure 31.1.2 shows the overall architecture of the proposed LLM accelerator with 4 ReRAM dies stacked on the logic die via 2048 face-to-face bumps for parallel read, delivering 25.6GB/s at 100MHz and 8MB memory capacity. The logic die mainly consists of a top controller, an MCU, a 64KB ISA buffer, 4 PLLs, an interconnect bus, a 1MB weight buffer, a 2MB global token buffer, an EMA controller (EMAC), an inter-chip transceiver, an LRU, a workload-decoupled out-of-order scheduler (WDOS), and an RS-PNM that includes a codebook fetcher unit (CFU), a ReRAM load interface (RLI), a tile-fused tensor engine (TFTE), and a non-linear processing unit (NLPU). During SD, the LRU performs local rotation for TLM, where the token allocator (TAU) sends upper/lower features to the reconfigurable FWHT array (RFA) or Hadamard accumulator unit (HAU) for decomposed FWHT. The rotated token is dynamically quantized with scaling factors bypassed to the TFTE for subsequent layer quantization. The RS-PNM utilizes the high bandwidth to load DLM CBs in ReRAM via RLI and fuses the token tiles that share the same CB entry by TFTE to avoid redundant CB loading. WDOS enables APSD by decoupling workloads into 4 parallel instruction queues, which are scheduled in an out-of-order manner with dependency-aware synchronization to maximize resource and bandwidth utilization.

Figure 31.1.3 illustrates the LRU with decomposed FWHT for low-bit outlier-free TLM quantization. The Hadamard matrix in the FWHT is built via a Kronecker product for power-of-two dimensions. However, a TLM often contains non-power-of-two (npot) channels. To handle them, the npot dimension n is typically factorized into $2^k \times m$ dimensions (e.g., $14336=2^9\times28$ for the LLaMA3-8B down_proj layer), where m is the size of a pre-computed npot Hadamard matrix [15]. Such factorization results in a cascaded FWHT-GEMM array, which requires numerous high-precision operators, leading to significant area overhead. To address this, we utilize npot Hadamard construction to limit the FWHT depth from 9 to a low-cost 6, and approximate the global rotation using overlapped upper and lower rotations with searched with (m, k) pairs such that their combined coverage spans the original dimension n. Then, LRU starts two-stage local rotation with (m, k). In each stage, the TAU assigns the token tiles to RFAs first, which are reconfigurable to support $2^1$-$2^6$ FWHT. To reduce the need for high-precision adders, adjacent FWHTs at early stages are merged to share inputs, and a lightweight router network further dispatches desired data based on the selected mode. Subsequently, the TAU allocates RFA outputs and the binary part of npot Hadamard tiles to the HAU for MAC-free accumulation by fusing FP16 Hadamard and the dynamic quantizers' scales. The LRU enables accurate W4A8 TLM quantization, achieving 3.82-to-3.93× speedup over BF16 SD while saving 92.7% area compared to global rotation.

Figure 31.1.4 depicts the RS-PNM architecture with the proposed BVQ algorithm. Unlike traditional vector quantization (VQ) methods [16-17] that incur heavy area overheads from index buffers and multi-port decoders [18]. BVQ performs block-level clustering by jointly learning blockwise CBs with INT4 QAT and block indices with Gumbel softmax reparameterization inspired by [19-20], which only requires a lightweight ISA decoder to retrieve the CBs. In RS-PNM, the MCU stores the DLM CBs into stacked ReRAM dies via SPI interface, after which the CFU triggers the ReRAM controller and RLI to load CBs from ReRAM to weight buffer. The RLI uses a double clock rate (200MHz) to stabilize the read data before transferring it to asynchronous FIFO groups for reliable clock domain crossing. To avoid data congestion caused by horizontal CB mapping at limited clock frequency, vertical CB mapping is employed. In addition, block dimensions within each CB are constrained by the per-die ReRAM bank width to maximize bandwidth utilization. However, vertical mapping results in redundant CB access issues when the complete weight is reconstructed on-chip, leading to extra latency overhead. To mitigate this, the tile fusion unit (TFU) fuses token tiles that share the same CB entry, ensuring each CB is fetched only once and thereby halving CB read latency. Moreover, TFUs facilitate both intra- and inter-layer parallelism by performing token fusion independently. Compared to W4A8 SD with LRU, the RS-PNM with INT4 BVQ achieves 1.1-to-1.46× speedup.

Figure 31.1.5 shows the APSD scheme with WDOS. A recent parallel SD method [14] enhances vanilla SD by inter-chip parallel draft-and-verify, which isolates DLM drafting and TLM verification workloads to avoid resource competition issues in intra-chip parallelism. However, the slow TLM verification leads to severe DLM idle time and wastes ReRAM bandwidth. The inter-chip parallelism further lowers both the ReRAM and DRAM bandwidth utilization as the memory interface in each chip cannot be shared across workloads. In contrast, APSD adaptively switches between short DL drafting and long DL parallel draft-and-verify based on whether previous draft tokens are all accepted and the first draft token matches TLM output, which alleviates DLM idleness. Moreover, APSD employs intra-chip parallelism enabled by WDOS with CB-interleaved intra-layer mapping (CILM). The CILM evenly allocates intra-layer CBs across chips and interleaves them within each DLM block to ensure full ReRAM bandwidth utilization at loading. Moreover, APSD workloads are simplified and decoupled into four instruction queues: inter-chip transceiver, compute, ReRAM load, and EMAC. Then, the intra-queue decoders extract dependency markers and send them to the inter-queue synchronizers, which jointly maintain a synchronous counter matrix for tracking readiness. An instruction is issued when its parent queues are ready, and its daughter queues are then notified. This dependency-aware scheduling efficiently enables intra-chip parallel draft-and-verify with high resource and bandwidth utilization, achieving 1.1-to-1.29× speed up over RS-PNM with 10-to-14% rejected DLM latency reduction.

Figure 31.1.6 shows the measurement results of the LLM accelerator, fabricated in 55nm using bumping-based face-to-face ReRAM-on-logic stacking technology. Die photos, specifications, 4-chip system, SEM/TEM images, and a ReRAM resistance distribution curve are shown in Fig. 31.1.7. The logic die operates at 63.5 to 285MHz at 0.89 to 1.40V, achieving 2.33TOPS peak performance. Each ReRAM die runs at 100MHz at 1.1V and consumes 49.54mW. The chip achieves 4.46-to-7.17× speedup and 3.74-to-4.85× energy savings over the BF16 SD baseline across various TLM/DLM pairs. Compared to the latest works [3-7], our chip integrates 3.43MB SRAM and 8MB stacked ReRAM with 25.6GB/s bandwidth. In a 4-chip system, ReRAM further scales to 32MB with 102.4GB/s bandwidth, sufficient to store all DLM CBs. The LRU supports reliable W4A8 quantization, achieving perplexity comparable to a SOTA W8A16 LLM accelerator [4] and better than works [6-7] that leverage 4b or sub-4b weight quantization. While most prior works [3–6] focus on compute-bound LLM prefilling optimization, the critical bottleneck lies in decoding. For a fair system-level comparison, we enhance each work with an LPDDR3 interface [21] to include EMA and 4-chip parallelism. As for LLaMA2-7B on the MT-Bench dataset that contains both prefilling and decoding, our chip achieves high decoding throughput of 17.82tokens/s and low energy consumption of 123.41mJ/token.

*Acknowledgement:*
This research was supported by ACCESS – AI Chip Center for Emerging Smart Systems, sponsored by InnoHK funding, Hong Kong SAR. The corresponding authors of this paper are Kwang-Ting Cheng (timcheng@ust.hk) and Liang Zhao (lzhao2020@zju.edu.cn).

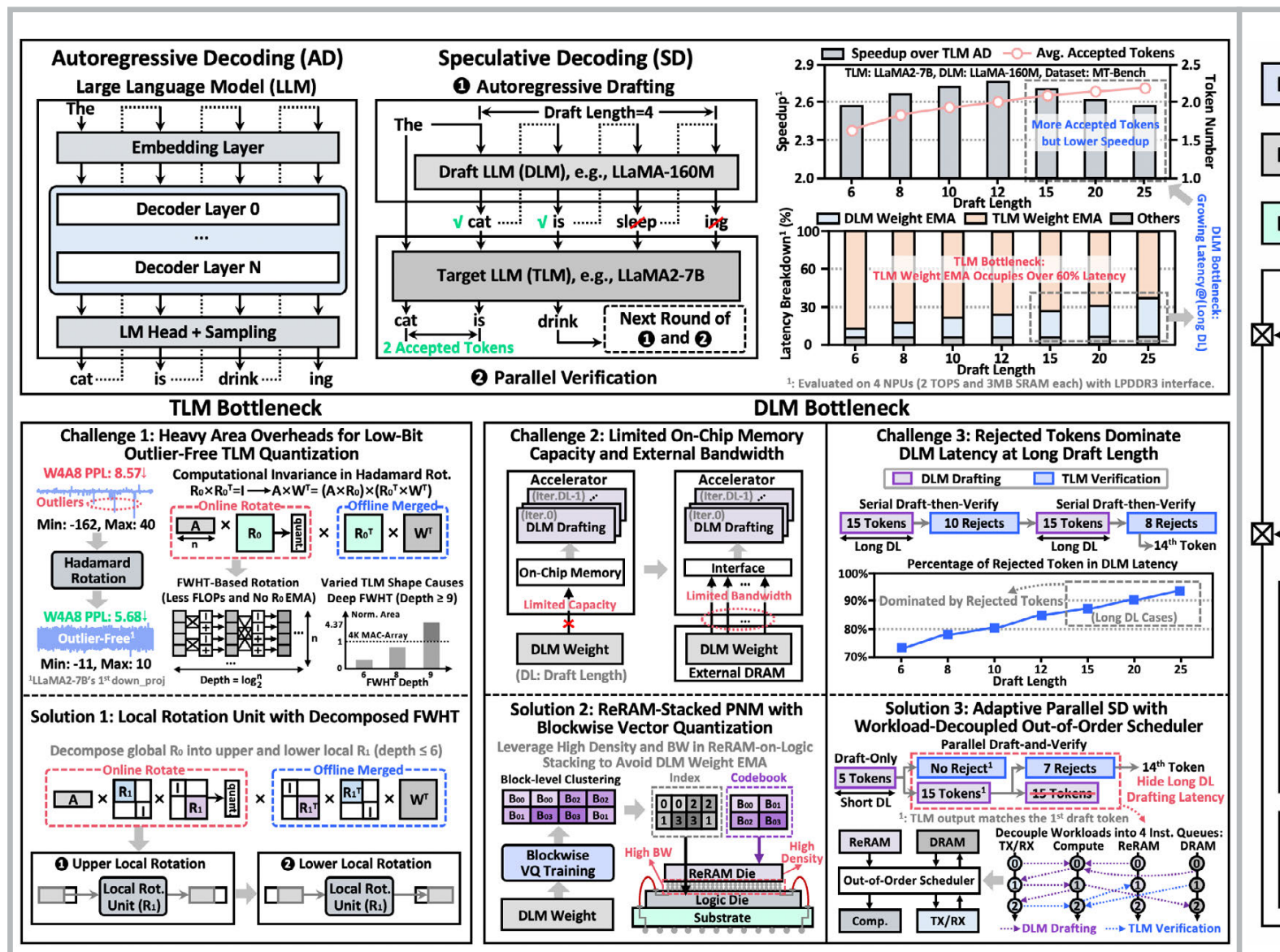

**Figure 31.1.1: Challenges raised by target and draft large language model (LLM) in speculative decoding (SD) and proposed solutions.**

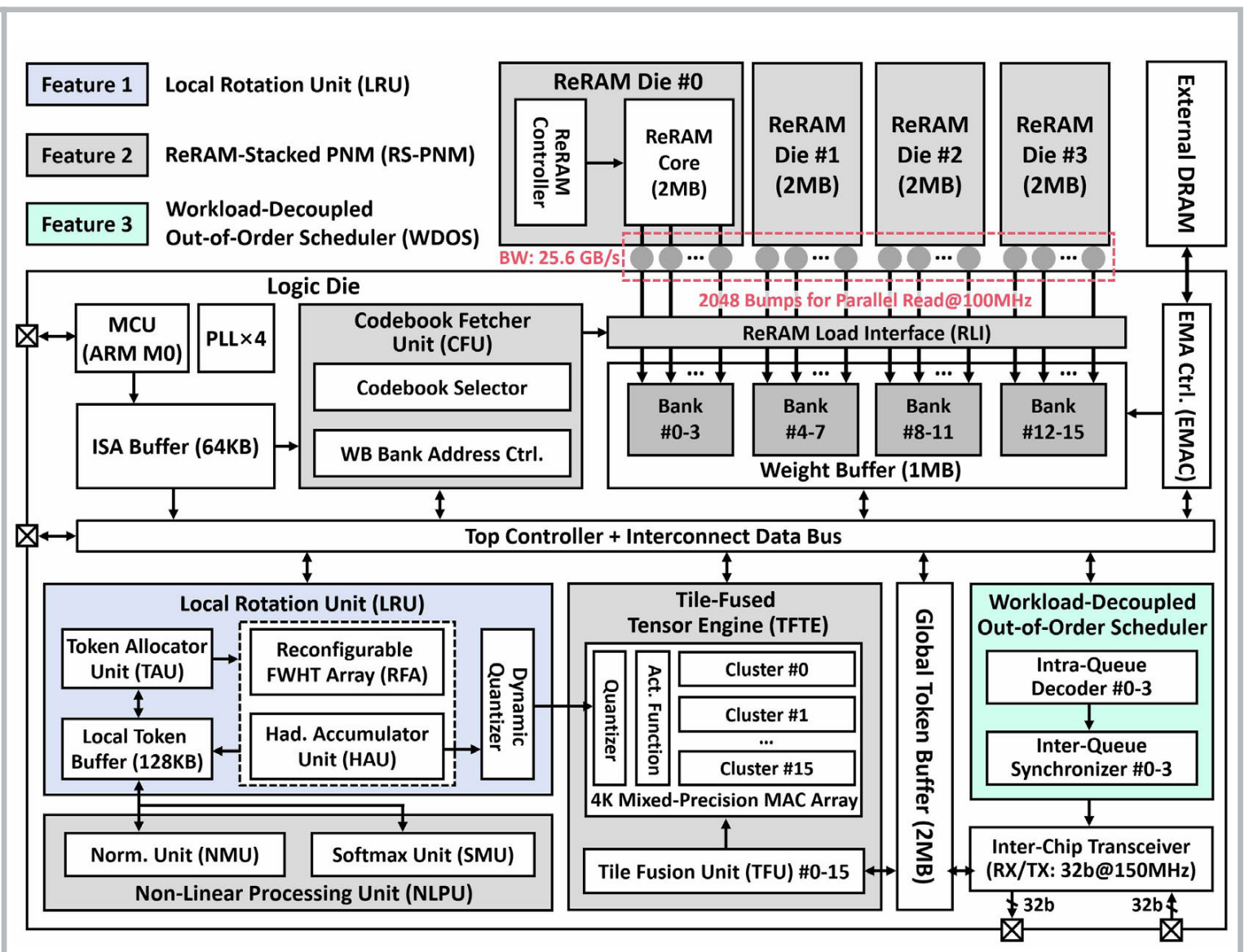

**Figure 31.1.2: Overall architecture and three main features of the LLM accelerator with bumping-based ReRAM die on logic wafer face-to-face stacking technology.**

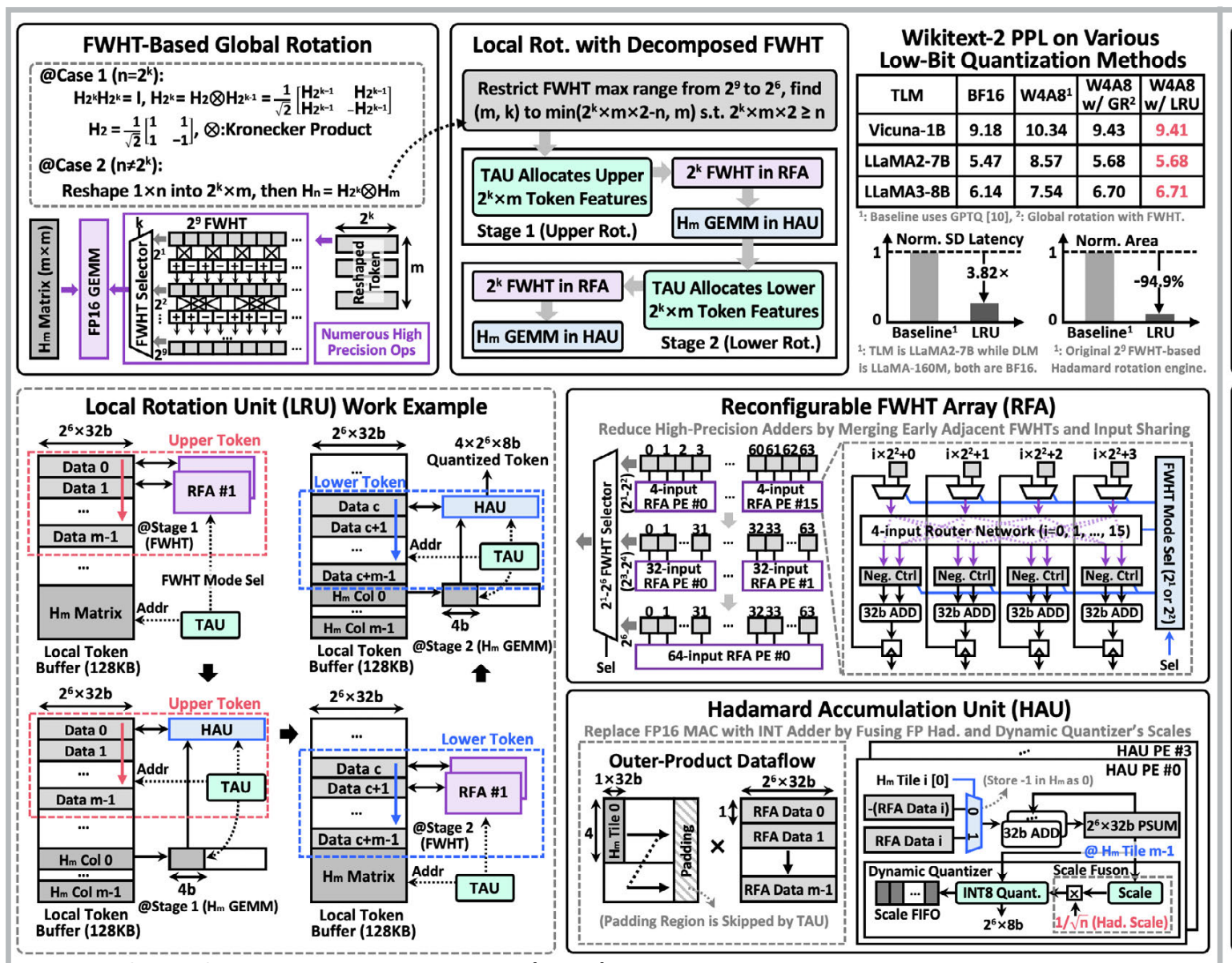

Figure 31.1.3: Local rotation unit (LRU) with proposed decomposed Fast Walsh-Hadamard Transform (FWHT) for outlier-free low-bit target LLM quantization.

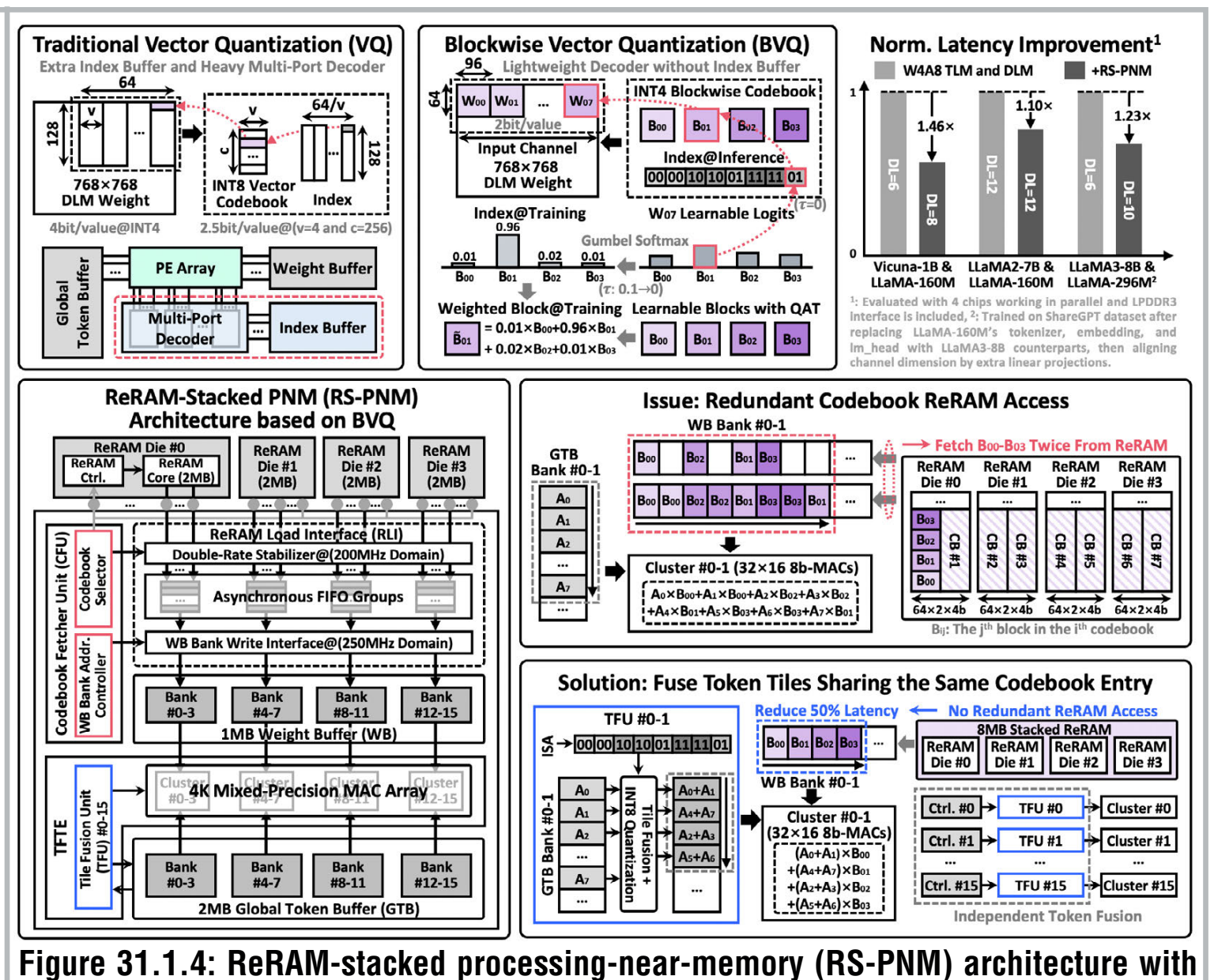

Figure 31.1.4: ReRAM-stacked processing-near-memory (RS-PNM) architecture with blockwise vector quantization (BVQ) to avoid draft LLM external memory access (EMA).

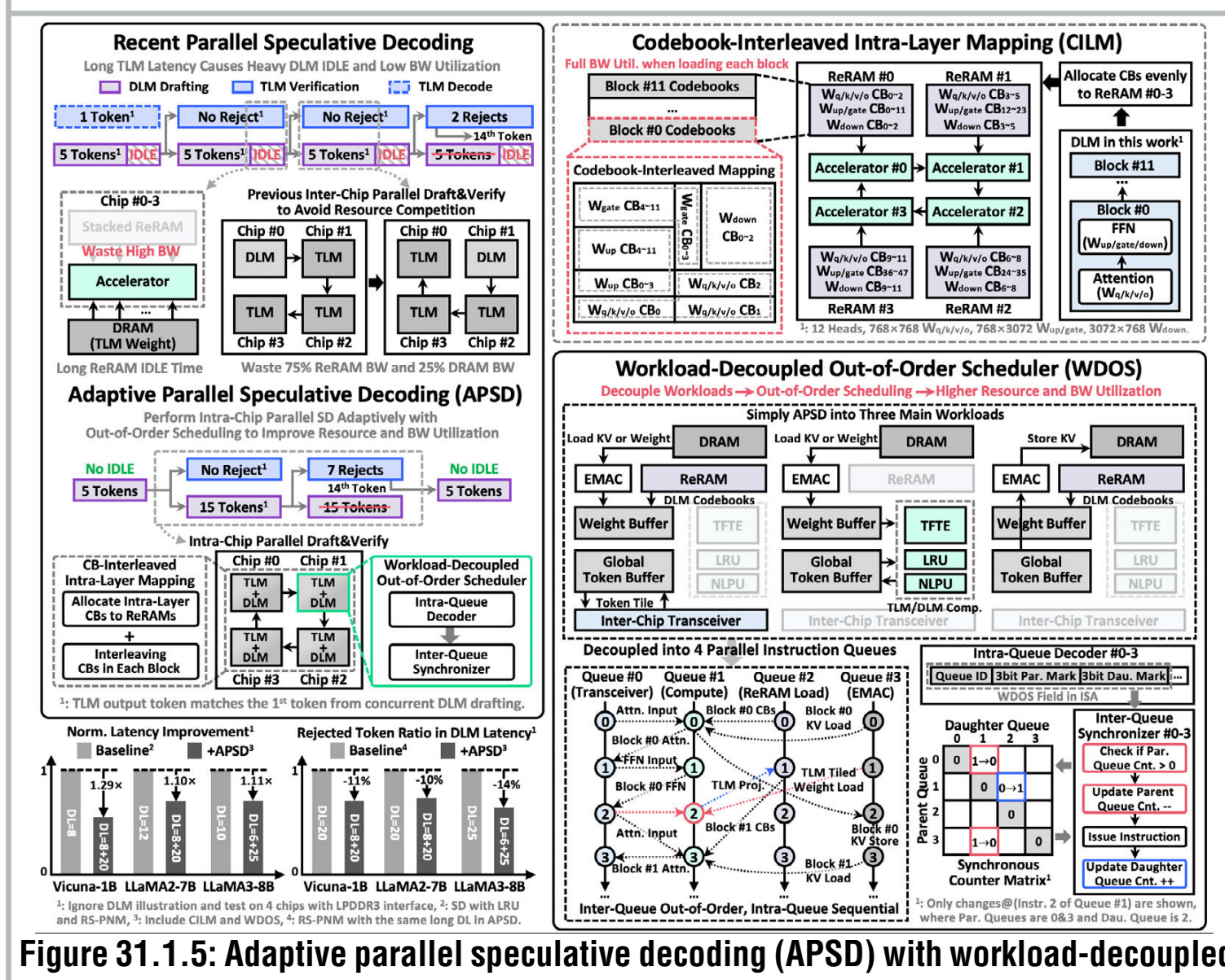

Figure 31.1.5: Adaptive parallel speculative decoding (APSD) with workload-decoupled out-of-order scheduler (WDOS) to achieve intra-chip parallel draft-and-verify with high resource utilization.

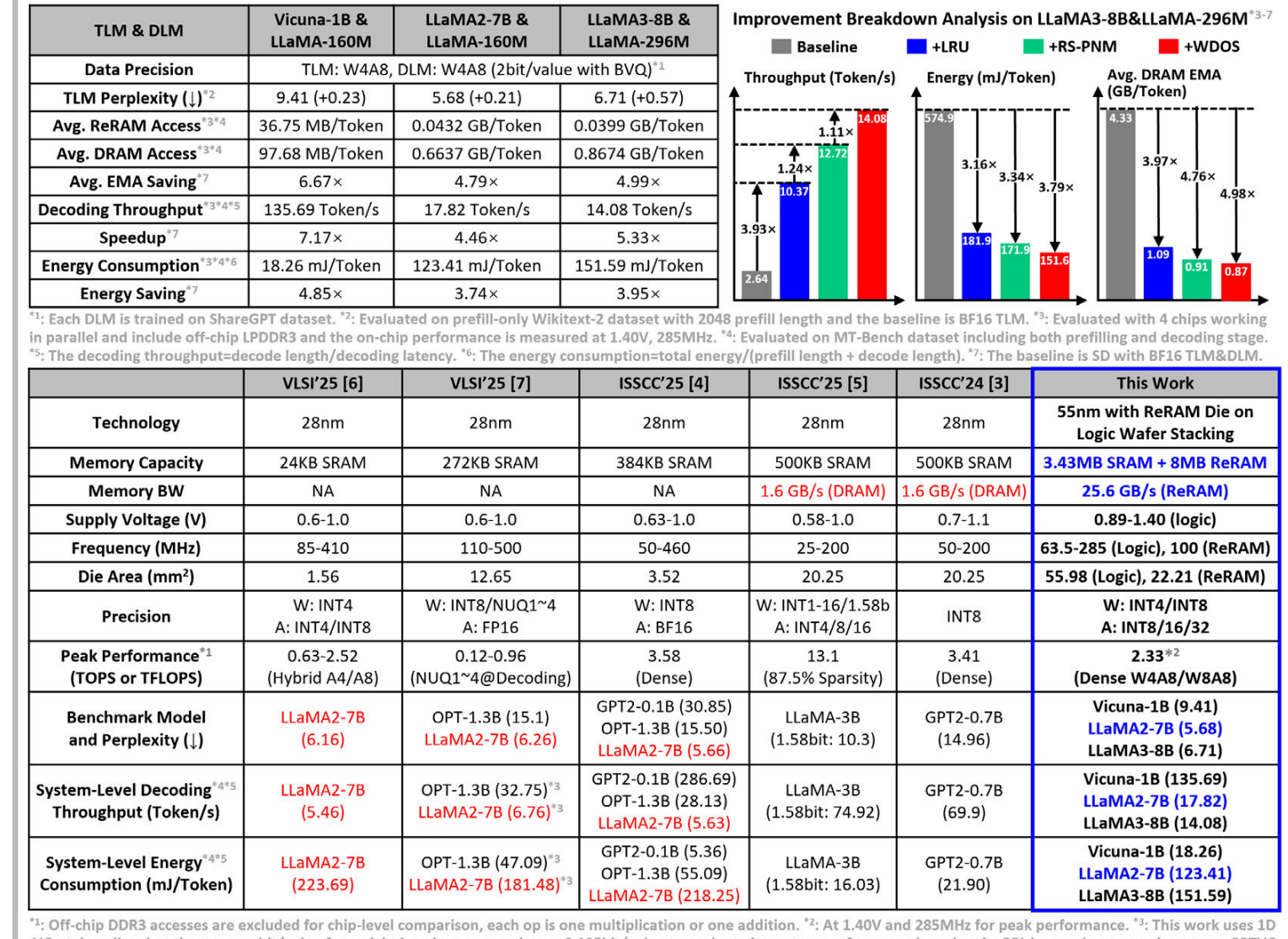

| TLM & DLM | Vicuna-1B & LLaMA-160M | LLaMA2-7B & LLaMA-160M | LLaMA3-8B & LLaMA-296M |
|---|---|---|---|
| Data Precision | TLM: W4A8, DLM: W4A8 (2bit/value with BVQ)*1 | | |
| TLM Perplexity (↓)*2 | 9.41 (+0.23) | 5.68 (+0.21) | 6.71 (+0.57) |
| Avg. ReRAM Access*3*4 | 36.75 MB/Token | 0.0432 GB/Token | 0.0399 GB/Token |
| Avg. DRAM Access*3*4 | 97.68 MB/Token | 0.6637 GB/Token | 0.8674 GB/Token |
| Avg. EMA Saving*7 | 6.67× | 4.79× | 4.99× |
| Decoding Throughput*3*4*5 | 135.69 Token/s | 17.82 Token/s | 14.08 Token/s |
| Speedup*7 | 7.17× | 4.46× | 5.33× |
| Energy Consumption*3*4*6 | 18.26 mJ/Token | 123.41 mJ/Token | 151.59 mJ/Token |
| Energy Saving*7 | 4.85× | 3.74× | 3.95× |

*1: Each DLM is trained on ShareGPT dataset. *2: Evaluated on prefill-only Wikitext-2 dataset with 2048 prefill length and the baseline is BF16 TLM. *3: Evaluated with 4 chips working in parallel and include off-chip LPDDR3 and the on-chip performance is measured at 1.40V, 285MHz. *4: Evaluated on MT-Bench dataset including both prefilling and decoding stage. *5: The decoding throughput=decode length/decoding latency. *6: The energy consumption=total energy/(prefill length + decode length). *7: The baseline is SD with BF16 TLM&DLM.

| | VLSI'25 [6] | VLSI'25 [7] | ISSCC'25 [4] | ISSCC'25 [5] | ISSCC'24 [3] | This Work |
|---|---|---|---|---|---|---|
| Technology | 28nm | 28nm | 28nm | 28nm | 28nm | 55nm with ReRAM Die on Logic Wafer Stacking |
| Memory Capacity | 24KB SRAM | 272KB SRAM | 384KB SRAM | 500KB SRAM | 500KB SRAM | 3.43MB SRAM + 8MB ReRAM |
| Memory BW | NA | NA | NA | 1.6 GB/s (DRAM) | 1.6 GB/s (DRAM) | 25.6 GB/s (ReRAM) |
| Supply Voltage (V) | 0.6-1.0 | 0.6-1.0 | 0.63-1.0 | 0.58-1.0 | 0.7-1.1 | 0.89-1.40 (logic) |
| Frequency (MHz) | 85-410 | 110-500 | 50-460 | 25-200 | 50-200 | 63.5-285 (Logic), 100 (ReRAM) |
| Die Area (mm²) | 1.56 | 12.65 | 3.52 | 20.25 | 20.25 | 55.98 (Logic), 22.21 (ReRAM) |
| Precision | W: INT4<br>A: INT4/INT8 | W: INT8/NUQ1~4<br>A: FP16 | W: INT8<br>A: BF16 | W: INT1-16/1.58b<br>A: INT4/8/16 | INT8 | W: INT4/INT8<br>A: INT8/16/32 |
| Peak Performance*1 (TOPS or TFLOPS) | 0.63-2.52 (Hybrid A4/A8) | 0.12-0.96 (NUQ1~4@Decoding) | 3.58 (Dense) | 13.1 (87.5% Sparsity) | 3.41 (Dense) | 2.33*2 (Dense W4A8/W8A8) |
| Benchmark Model and Perplexity (↓) | LLaMA2-7B (6.16) | OPT-1.3B (15.1)<br>LLaMA2-7B (6.26) | GPT2-0.1B (30.85)<br>OPT-1.3B (15.50)<br>LLaMA2-7B (5.66) | LLaMA-3B (1.58bit: 10.3) | GPT2-0.7B (14.96) | Vicuna-1B (9.41)<br>LLaMA2-7B (5.68)<br>LLaMA3-8B (6.71) |
| System-Level Decoding*4*5 Throughput (Token/s) | LLaMA2-7B (5.46) | OPT-1.3B (32.75)*3<br>LLaMA2-7B (6.76)*3 | GPT2-0.1B (286.69)<br>OPT-1.3B (28.13)<br>LLaMA2-7B (5.63) | LLaMA-3B (1.58bit: 74.92) | GPT2-0.7B (69.9) | Vicuna-1B (135.69)<br>LLaMA2-7B (17.82)<br>LLaMA3-8B (14.08) |
| System-Level Energy*4*5 Consumption (mJ/Token) | LLaMA2-7B (223.69) | OPT-1.3B (47.09)*3<br>LLaMA2-7B (181.48)*3 | GPT2-0.1B (5.36)<br>OPT-1.3B (55.09)<br>LLaMA2-7B (218.25) | LLaMA-3B (1.58bit: 16.03) | GPT2-0.7B (21.90) | Vicuna-1B (18.26)<br>LLaMA2-7B (123.41)<br>LLaMA3-8B (151.59) |

*1: Off-chip DDR3 accesses are excluded for chip-level comparison, each op is one multiplication or one addition. *2: At 1.40V and 285MHz for peak performance. *3: This work uses 1D VQ at decoding, but the average bit/value for weight is unknown, we choose 3.125bit/value to evaluate its system performance based on its PPL loss and our experiments on GPTVQ [16]. *4: Evaluated at the peak performance on MT-Bench dataset with an average 454 prefill length and 512 decode length across all LLMs. Using 4-chip parallelism and each features a LPDDR3 interface (6.4GB/s and 75% bandwidth utilization) with a 120pJ/Byte energy cost following [21]. *5: The weight of LM_Head is INT4 while activation uses each work's setting.

Figure 31.1.6: Measurement results and comparison with state-of-the-art LLM accelerators.

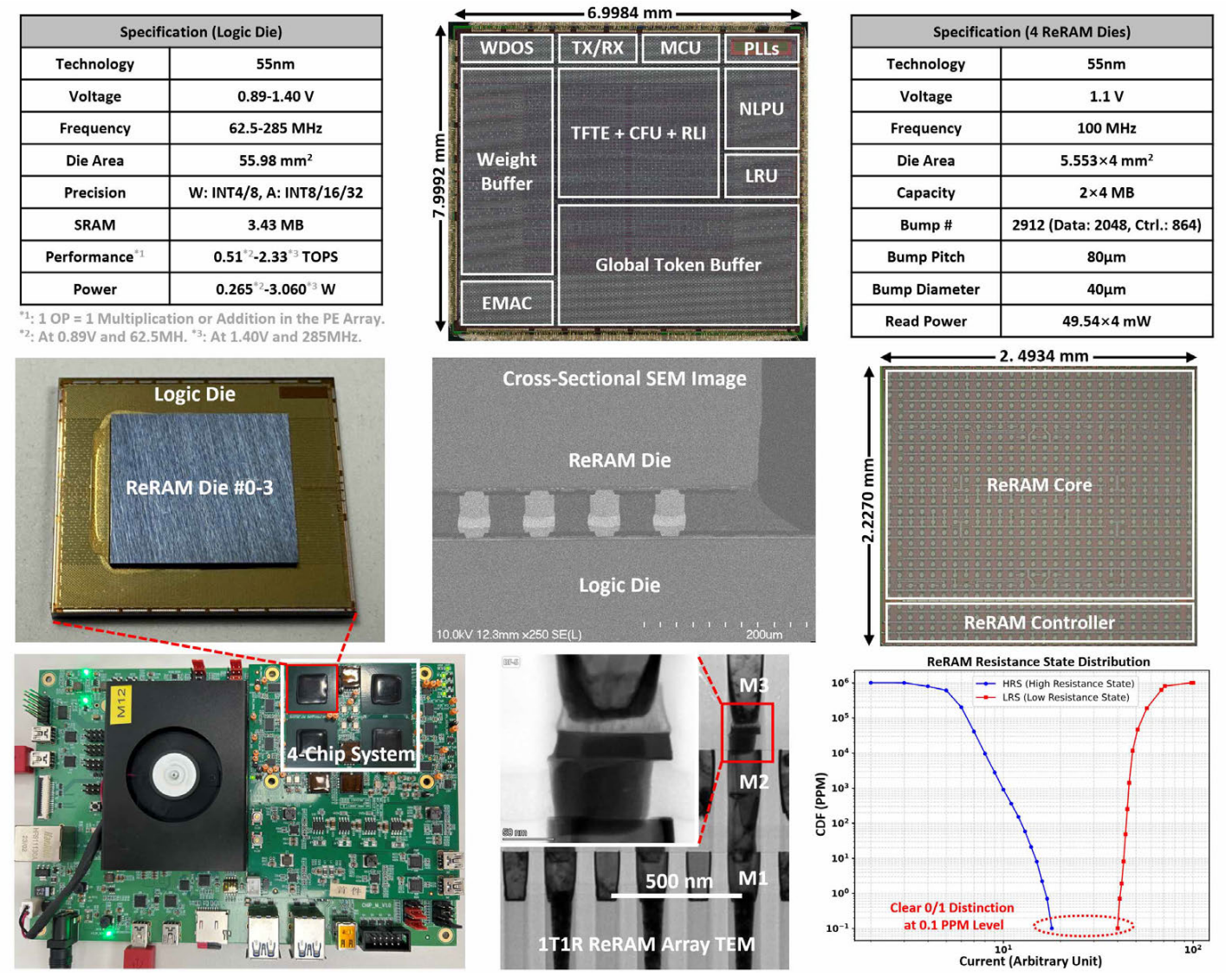

| Specification (Logic Die) | |
|---|---|
| Technology | 55nm |
| Voltage | 0.89-1.40 V |
| Frequency | 62.5-285 MHz |
| Die Area | 55.98 mm² |
| Precision | W: INT4/8, A: INT8/16/32 |
| SRAM | 3.43 MB |
| Performance*1 | 0.51*2-2.33*3 TOPS |
| Power | 0.265*2-3.060*3 W |

*1: 1 OP = 1 Multiplication or Addition in the PE Array. *2: At 0.89V and 62.5MH. *3: At 1.40V and 285MHz.

| Specification (4 ReRAM Dies) | |
|---|---|
| Technology | 55nm |
| Voltage | 1.1 V |
| Frequency | 100 MHz |
| Die Area | 5.553×4 mm² |
| Capacity | 2×4 MB |
| Bump # | 2912 (Data: 2048, Ctrl.: 864) |
| Bump Pitch | 80µm |
| Bump Diameter | 40µm |
| Read Power | 49.54×4 mW |

Figure 31.1.7: Die photos, specifications, 4-chip system, SEM/TEM images of the ReRAM-on-logic stacking interface/ReRAM chip, and ReRAM resistance distribution curve.